# Tackling climate change through energy efficiency: mathematical models for evidence-based public policy recommendations

F. Gallo[(1)], P. Contucci[(2)], A. Coutts[(3)], I. Gallo[(2)]
(1) Office of Climate Change, UK Government
(2) Dipartimento di Matematica, Universita' di Bologna
(3) Department of Politics and International Relations, University of Oxford

**Abstract**

**Promoting and increasing energy efficiency is a promising method of reducing $CO_2$ emissions and avoiding the potentially devastating effects of climate change. The question is: How do we induce a cultural or behavioural change whereby people nationally and globally adopt more energy efficient lifestyles?**

**We propose a new family of mathematical models, based on a statistical mechanics extension of discrete choice theory, that offer a set of formal tools to systematically analyse and quantify this problem. An application example could be to predict the percentage of people choosing to buy new energy efficient light bulbs instead of the traditional incandescent versions. Through statistical evaluation of survey responses, the models can identify the key driving factors in the decision-making process; for example, the extent to which people imitate each other. These models allow us to incorporate the effect of social interactions could help us identify 'tipping points' at a societal level. This knowledge could be used to trigger structural changes in our society. The results may provide tangible and deliverable evidence-based policy options to decision-makers.**

**We believe that these models offer an opportunity for the research community–in both the social and physical sciences–and decision-makers in the private and public sectors to work together towards preventing the potentially devastating social, economic and environmental effects of climate change.**

## 1. Introduction

Climate change is one of the greatest environmental, social and political challenges facing humankind. In order to prevent it, we need to significantly reduce global $CO_2$ and greenhouse gas emissions over the next few decades. Energy efficiency offers an option to achieve this and involves a number of social and economic advantages. For example, it does not require us to reduce our standard of living and could result in significant financial savings.

Therefore, we are faced with an enormous problem: Can we achieve a cultural, behavioural and structural change–both nationally and internationally–whereby people decide out of their own volition to adopt more energy efficient behaviours and lifestyles? The purpose of this paper is to present a tool that could help decision-makers achieve this goal; in particular, to induce our society to make a transition towards a sustainable way of life.

For the sake of analysis, one may view this issue as a binary choice problem: nationally or globally a population may choose either to continue an energy inefficient lifestyle or replace it with an energy efficient one. One key difficulty with this problem statement is that it is too vague and general. However, we can overcome this difficulty by breaking the problem into smaller, more tractable, components that can be clearly specified, quantified and attacked.

Figure 1 shows a schematic illustration of this approach. As an example, we could focus on energy efficiency measures and view them individually as binary choices: for instance buying new energy efficient light bulbs instead of energy guzzling incandescent bulbs. In this way, the apparently intractable problem of shifting our society's behaviour from an energy inefficient lifestyle to a sustainable one is reduced to simpler problems such as inducing people in *specific geographical areas* and from *specific socio-economic backgrounds* to change *specific choices*, such as buying energy efficient bulbs.

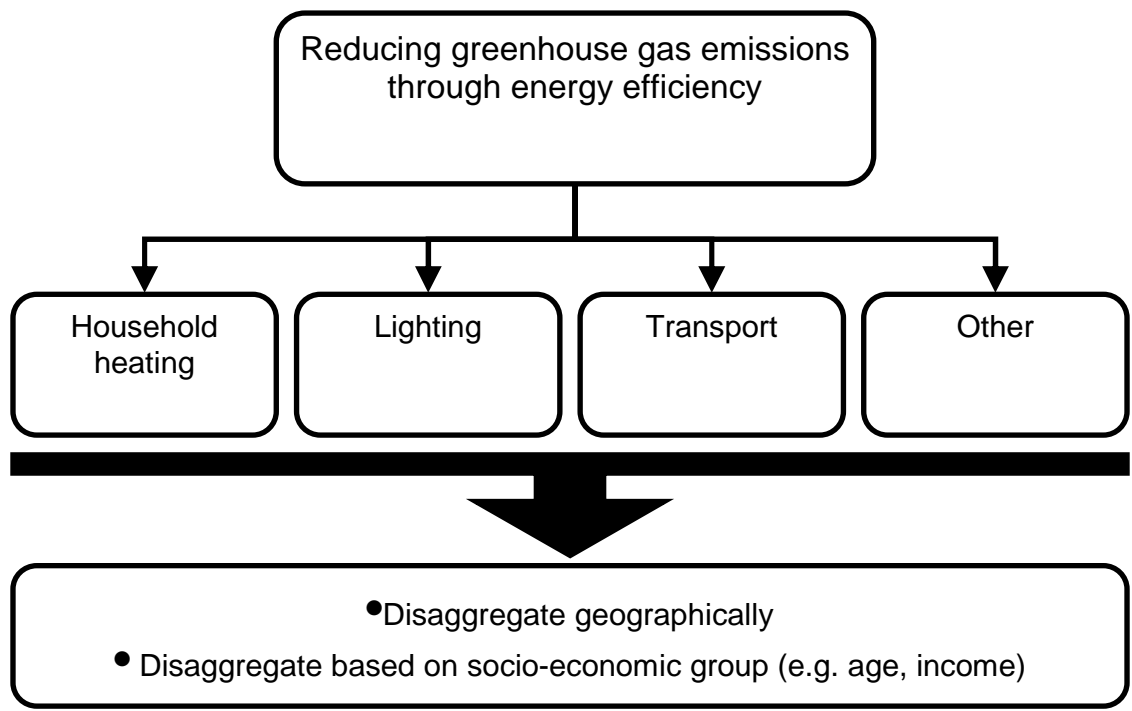

**Figure 1:** *Schematic illustration of how the goal of achieving a global culture change may be broken into smaller components that are easier to analyse. The models presented in this paper offer a rigorous, quantitative methodology to obtain the evidence to inform policy-makers.*

The purpose of this paper is to introduce a new multidisciplinary 'family' of models that may help policy-makers to tackle the above problem. In particular, these models may help decision-makers adopt policy options that will induce widespread behaviour change, and perhaps even structural or cultural change. To this end, this family of models offers two specific tools to decision-makers:

1. First, *a framework to systematically deconstruct this apparently intractable problem into smaller, more manageable pieces*, as shown in Figure 1. This would allow researchers to focus on and analyse one specific problem at a time, find the most relevant policy options, and then move on to the next problem. It is important to note that this is also possible because many choices are generally independent of each other: for example, the choice of buying energy-efficient light bulbs may be assumed to be independent of the mode of transport used to go to work.

2. Second, these models offer a *rigorous and quantitative bottom-up tool* to identify and understand the key incentives that drive people's decisions. As will be shown in the following sections, the end product of this analysis is a formula known as a utility function that describes people's preferences in a given context. This formula can be designed so that it includes variables, or policy-levers, that decisions-makers can manipulate in order to induce behaviour change. An example of such a variable would be the level of taxation on a given consumer product, say energy efficient light bulbs. As will be explained later in more detail, this utility function is obtained empirically from data.

The remainder of this paper is structured as follows. *Section 2* introduces the behavioural models. *Section 3* provides details of energy efficiency in the context of climate change and gives further evidence of the benefits of using the above behavioural models in this context. *Section 4* briefly discusses how these models could be applied to many areas of social policy. Finally, *Section 5* concludes and proposes some ideas for further research.

## 2. Behavioural models: beyond "Rational Man" and *Homo Economicus*

This section presents a family of behavioural models that could be applied directly to the problem of increasing energy efficiency through behaviour change. These models have been developed to overcome the main limitations of the *Homo Economicus* or 'Rational Man' model (Persky 1995), including:

- access to limited information and emotions
- social interactions and imitation

The next subsection gives an overview of discrete choice theory, which is an econometric tool that has been used for over three decades to understand people's preferences in issues ranging from transport to healthcare. This theory not only describes the rational aspects of human choice, but can also account empirically for factors such as emotions or imperfect information.

## 2.1 Bounded rationality and emotion: Discrete Choice Analysis

Discrete choice analysis is a well-established research tool that has been applied to real social phenomena for more than thirty years. Due to the development of this theory, Daniel McFadden was awarded the Nobel Prize in Economics in 2000, for bringing economics closer to quantitative scientific measurement. Figure 2 shows an example where the model prediction was 98 percent accurate.

**Table 1. Prediction Success Table, Journey-to-Work**
(Pre-BART Model and Post-BART Choices)

| Cell Counts | Predicted Choices | | | | |
|---|---|---|---|---|---|
| Actual Choices | Auto Alone | Carpool | Bus | BART | Total |
| Auto Alone | 255.1 | 79.1 | 28.5 | 15.2 | 378 |
| Carpool | 74.7 | 37.7 | 15.7 | 8.9 | 137 |
| Bus | 12.8 | 16.5 | 42.9 | 4.7 | 77 |
| BART | 9.8 | 11.1 | 6.9 | 11.2 | 39 |
| Total | 352.4 | 144.5 | 94.0 | 40.0 | 631 |
| | | | | | |
| Predicted Share | 55.8% | 22.9% | 14.9% | 6.3% | |
| (Std. Error) | (11.4%) | (10.7%) | (3.7%) | (2.5%) | |
| Actual Share | 59.9% | 21.7% | 12.2% | 6.2% | |

**Figure 2:** *Discrete predictions against actual use of travel modes in San Francisco, 1975 (source: McFadden 2001)*

The purpose of discrete choice theory is to describe people's behaviour. It is an econometric technique to infer people's preferences from empirical data. In discrete choice theory the decision-maker is assumed to make choices that maximise his/her own benefit. Their 'benefit' is described by a mathematical formula, a utility function, which is derived from data collected in surveys. This utility function includes rational preferences, but also accounts for elements that deviate from rational behaviour.

Discrete choice models, however, do not account for 'peer pressure' or 'herding effects': individual decisions are assumed to be driven by influences such as prices for goods and overall quality of services. In other words, discrete choice assumes that people's decisions are unaffected by the choices made by other people. We shall see in Section 2.2, however, that there are good reasons–both empirical and theoretical–to believe that influences of other people might play a crucial and quantifiable role in the overall behaviour of society.

It is nonetheless a fact that the standard performance of discrete choice models is close to optimal for the analysis of many phenomena where peer influence is perhaps not a major factor in an individual's decision: Figure 2 shows an example of this. The table (taken from McFadden 2001) compares predictions and actual data concerning the use of travel modes, before and after the introduction of new rail transport system called BART in San Francisco, 1975. We see a remarkable agreement between the predicted share of people using BART (6.3%), and the actual measured figure after the introduction of the service (6.2%).

## 2.1.1 Theory

In discrete choice theory each decision process is described mathematically by a utility function, which each individual seeks to maximize. As an example, a binary choice could be either to cycle to work or to catch a bus. The utility function for choosing the bus may be written as:

$$U = \sum_a \beta_a x_a + \sum_a \lambda_a y_a + \varepsilon.$$
Eq. 1

The variables $x_a$ are attributes that describe the alternatives, for example an attribute could be the bus fare or the journey time. On the other hand, the $y_a$ are socio-economic variables that define the decision-maker, for example their age, gender or income. It is this latter set of parameters that allows us to zoom in on specific geographical areas or socio-economic groups.

The $\beta_a$ and $\lambda_a$ are parameters that need to be estimated empirically using survey data. The key property of these parameters is that they quantify the relative importance of any given attribute in a person's decision: the larger the parameter's value, the more it will affect a person's choice. For example, we may find that certain people are more affected by the journey time than the bus fare; therefore changing the fare may not influence their behaviour significantly. The next section will explain how the value of these parameters is estimated from empirical data.

It is an observed fact (Luce and Suppes 1965, Ariely 2008) that choices are not always perfectly rational. For example, someone who usually goes to work by bus may one day decide to cycle instead. This may be because it was a sunny day, or because they wanted to exercise. This unpredictable component of people's choices is accounted for by the random term $\varepsilon$.

The functional distribution of $\varepsilon$ can take different forms. This gives rise to different possible models. For instance, if $\varepsilon$ is assumed be extreme-value distributed, the resulting model is called a *logit* model (Ben-Akiva and Lerman 1985). Logit models are widely used by discrete choice practitioners. A very powerful property of logit models is that they admit a closed form solution for the probability of choosing a particular alternative, say catching a bus rather than cycling to work :

$$P = \frac{e^V}{1+e^V},$$
Eq. 3

where $V$ is the deterministic part of the utility $U$ in Eq. 1:

$$V = \sum_a \beta_a x_a + \sum_a \lambda_a y_a.$$

In words, this describes the rational preferences of the decision maker.

As will be explained later on, Eq. 3 is analogous to the equation–in statistical mechanics–that describes the equilibrium state of a perfect gas of heterogeneous molecules: just like gas molecules react to external forces differently depending, for instance, on their mass and charge, discrete choice describes individuals as experiencing heterogeneous influences in their decision-making, according to their own socio-economic attributes, such as gender and wealth. A question arises spontaneously: do people and gases behave in the same way? The answer is that in some circumstances they might. Models are idealisations of reality. Eq. 3 is telling us that the same equation may describe idealised aspects of both human and gas behaviour; in particular, how individual behaviour relates to macroscopic or societal variables. These issues go beyond the scope of this paper, but it is important to note that Eq. 3 offers a mathematical and intuitive link between econometrics and physics. The importance of this 'lucky coincidence' cannot be overstated, and some of the implications will be discussed later on in more detail.

**2.1.2 Empirical estimation**

Discrete choice may be seen as a purely empirical model. The utility function given by Eq. 1 is very general: it may be seen as describing the decision process of a typical human being. In order to specify the actual functional form associated with a specific group of people facing a specific choice, empirical data is needed. The actual utility function is then specified by estimating the numerical values of the parameters $\beta_a$ and $\lambda_a$. As mentioned earlier, these parameters quantify the relative importance of the attribute variables $x_a$ and $y_a$. For example, costs are almost always associated with negative parameters: this means that the higher the price of an alternative, the less likely people will be to choose it. This makes intuitive sense. What discrete choice offers is a quantification of this effect.

There are two types of data that may be used to estimate the values of the parameters $\beta_a$ and $\lambda_a$:

1. revealed preference data
2. stated preference data

Revealed preference refers to choices that people have made in the past. For example, 'roadside interviews' collect information about people's actual travel choices, including the chosen route, time of day and the mode of transport. On the other hand, stated preference data is based on hypothetical questions. For example, businesses may be interested to learn about people's preferences in view of the imminent launch of a new product. Once the data has been collected, the model parameters may be estimated by standard statistical techniques. In practice, Maximum Likelihood estimation methods are used most often (see, e.g., Ben-Akiva and Lerman (1985), chapter 4).

**2.1.3 Applications**

Discrete choice models have been used to study people's preferences since the nineteen seventies (McFadden 2001). Initial applications focused on transport (Train 2003, Ortuzar, J. and Wilumsen, L. 2001). These models have been used to develop national and regional transport models around the world, including in the UK, the Netherlands (Fox et al 2003), as well as Copenhagen (Paag 2001). Discrete choice modelling has also been applied to a range of societal problems such as healthcare (Gerard *et al* 2003; Ryan and Gerard 2003), telecommunications (Ida and Kuroda 2006) and social care (Ryan et al 2006). In particular, discrete choice is especially well suited to inform policy-making for a number of reasons. First, the fact that it is rooted in empirical data and that it has a rigorous and transparent methodology make it a trustworthy tool to use for evidence-based policy-making. Second, the utility functions that discrete choice models produce allow researchers to test concrete policy scenarios by changing variables such as the level of taxation.

## 2.2 Social interactions: Statistical Mechanics

This section presents a more recent extension of discrete choice theory that allows us to rigorously account for social interaction in human behaviour, including social norms and peer pressure. Based on well-established theories in mathematical physics, these models predict the existence of 'tipping points' and structural changes (see Figure 3). This could potentially be used to devise highly cost-effective social policies to induce cultural change as well as behaviour change at the individual level. For instance, we may find that a small subsidy on the costs of cavity wall insulation induces a 'critical mass' of people to change their behaviour. This in turn could hit a 'tipping point' whereby, through imitation, a large fraction of the population suddenly decides to insulate their own homes just because their societal peers seem to be doing so. In other words, insulating your house becomes 'fashionable'.

A key limitation of discrete choice theory is that it does not formally account for social interactions and imitation. In discrete choice theory each individual's decisions are based purely on personal preferences alone, and are not affected by other people's choices. However, there is a great deal of theoretical and empirical evidence to suggest that an individual's behaviour, attitude, identity and social decisions are influenced by that of others through vicarious experience or social influence, persuasions and sanctioning (Akerlof 1997; Bandura 1986). These theories specifically relate to the interpersonal social environment including social networks, social support, role models and mentoring. The key insight of these theories is that an individual's behaviour and decisions are affected by their relationships with those around them– e.g. their parents or their peers.

Mathematical models that take social influence into account have been considered by social psychology since the 1970s (see Scheinkman 2008 for a short review). In particular, influential works by Schelling (1978) and Granovetter (1978) have shown how models where individuals take into account the mean behaviour of others are capable of reproducing, at least qualitatively, the dramatic opinion shifts observed in real life (e.g. financial bubbles or street riots). In other words, they observed that the *interaction* built into their models was inextricably linked to the appearance of *structural changes* on a phenomenological level in the models themselves.

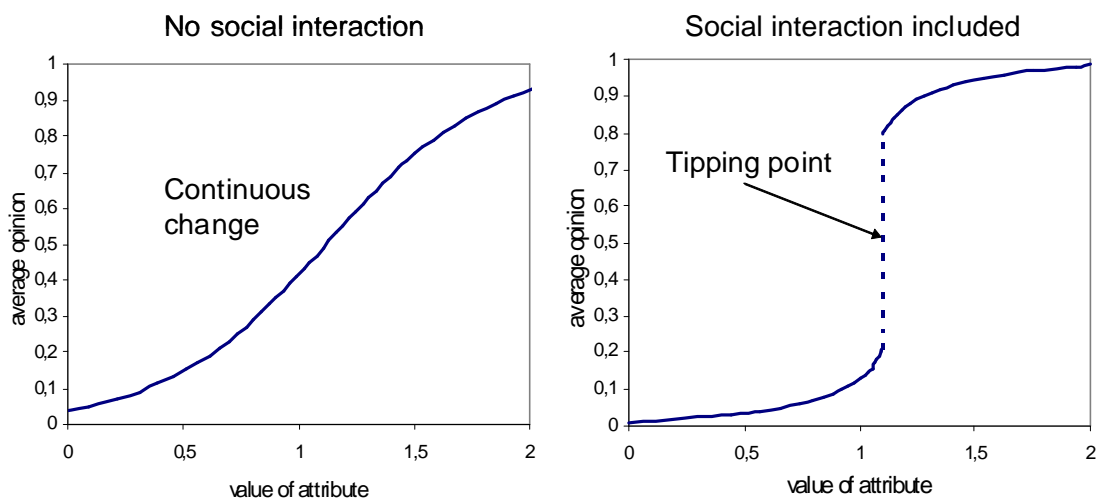

**Figure 3:** *The diagram illustrates how the inclusion of social interactions (right) leads to the existence tipping points. By contrast models that do not account for social interactions cannot account for the tipping points.*

Figure 3 illustrates the difference between models that include social interactions (right) and those that don't (left). In the former case, societal tipping points and structural cultural changes arise naturally. The vertical axis represents the average opinion across the population, for example buying efficient light bulbs or not. The horizontal axis shows the value of any one attribute, such as price or light quality. In the absence of social interactions (left diagram), the average opinion shifts gradually from one alternative to the other. However, if we account for social interactions (right), we see that attributes can reach a critical value at which we observe a tipping point or structural change. The model presented in this paper offers a rigorous tool to identify these tipping points.

This could have a significant impact on policy-making: it could potentially allow policy-makers to systematically identify policy options that could be very cost-effective: investing in a small change in the right attributes could lead to significant changes in society.

The research course initiated by Schelling was eventually linked to the development of the discrete choice analysis framework at the end of the nineties. These two models were merged by Brock and Durlauf (2001; 2005), who suggested a direct econometric implementation of models involving social interactions.. In order to accomplish this, Brock and Durlauf had to delve into the implications of a model where an individual takes into account the behaviour of others when making a discrete choice. This could only be done by considering a new utility function which depended on the choices of all other people.

This new utility function was built by starting from the assumptions of discrete choice analysis. The utility function reflects what an individual considers desirable: if we hold that people consider it desirable to conform to the ideas of people they interact with, it follows that an individual's utility increases when he agrees with other people (Bond and Smith 1996).

Symbolically, we can say that when an individual $i$ makes a choice, his/her utility for that choice increases by an amount $J_{ij}$ when another individual $j$ agrees with him/her, thus defining a set of interaction parameters $J_{ij}$ for all couples of individuals. The new utility function for individual $i$ hence takes the following form:

$$U_i = \sum_j J_{ij}\sigma_j + \sum_a \beta_a x_a^{(i)} + \sum_a \lambda_a y_a^{(i)} + \varepsilon \qquad \text{Eq. 4}$$

where the sum $\sum_j$ ranges over all individuals, and the symbol $\sigma_j$ is equal to 1 if $j$ agrees with $i$, and 0 otherwise.

Analysing the general case of such a model is a daunting task[1]. Fortunately, this problem has been considered by statistical mechanics since the end of the 19$^{th}$ century. Indeed, the first success of statistical mechanics was to give a microscopic explanation of the laws governing perfect gases, and this was achieved thanks to a formalism which is strictly equivalent to the one obtained by discrete choice analysis in Eq. 3.

The interest of statistical mechanics eventually shifted to problems concerning interaction between particles and–as daunting as the problem described by Eq. 4 may be–statistical physics has been able to identify some restrictions on models of this kind to make them tractable while still retaining great descriptive power, as shown in the work of Pierre Weiss (Weiss 1907) regarding the behaviour of magnets.

---

[1] The reason for this difficulty is that the choice of another individual $j$ is itself a random variable, which in turn correlates the choices of all individuals.

The simplest way devised by physics to deal with such a problem is called a *mean field* assumption, where interactions are assumed to be of a uniform and global kind. This leads to a manageable closed form solution, and a model that is consistent with the models of Schelling and Granovetter. Moreover, this model is also shown by Brock and Durlauf to be closely linked to the assumption of *rational expectations* from economic theory, which assumes that the observed behaviour of an individual must be consistent with his belief about the opinion of others.

By assuming *mean field* or *rational expectations* we can rewrite Eq. 4 in the tamer form

$$U_i = Jm + \sum_a \beta_a x_a^{(i)} + \sum_a \lambda_a y_a^{(i)} + \varepsilon \qquad \text{Eq. 5}$$

where *m* is the average opinion of a given individual, and this average value is coupled to the model parameters by a closed form formula.

If we now define $V_i$ to be the deterministic part of the utility, similarly as before,

$$V_i = Jm + \sum_a \beta_a x_a^{(i)} + \sum_a \lambda_a y_a^{(i)},$$

we have that the functional form of the choice probability, given by Eq. 3,

$$P_i = \frac{e^{V_i}}{1 + e^{V_i}},$$

remains unchanged, allowing the empirical framework of discrete choice analysis to be used to test the theory against real data. This sets the problem as one of heterogeneous interacting particles, and the physics of such mean-field systems has been shown to be analytically tractable (see, e.g., Contucci et al 2007).

The mean field assumption represents a first order approximation: it considers a kind of interaction that is fixed and uniform across the whole population. However, if more realism is needed, one should bear in mind that statistical physics has built throughout the twentieth century the expertise needed to consider a wide range of forms for the interaction parameters $J_{ij}$, of both deterministic and random nature. This means that a partial success in the application of mean field theory to social phenomena might be enhanced by browsing through a rich variety of well developed, though analytically more demanding, theories.

Nevertheless, an empirical attempt to assess the actual descriptive and predictive power of such models has not been carried out to date. The natural course for such a study would be to start by empirically testing the mean field picture, and then to proceed by enhancing it with the help available from the econometrics, social science, and statistical physics communities.

It is worthwhile to remark that the kind of cross-bred models considered in this paper offer a quantitative estimate of the role of social interactions in a decision making process. This is a relevant fact, since statistical mechanics tells us that a model involving social interactions is capable of exhibiting 'tipping point' like behaviour at a societal level.

These tipping points or structural changes, known as *phase transitions* in physics, cannot be predicted by standard discrete choice analysis, due to the regularity of the equations arising from it. On the other hand, it is a fact that sudden dramatic changes can be observed in the behaviour of large groups of people.

Therefore, by successfully implementing the kind of models considered here, researchers may gain the ability to study quantitatively a whole new range of human phenomena. As a consequence, policy-makers working in areas where the interaction between individuals could play a key-role, such as increasing energy efficiency of households, may acquire a valuable new tool.

## 3. Potential application - Energy efficiency and Climate Change

These types of models can be applied to any problem involving individuals making choices out of a finite set of alternatives. Applications over the past three decades include transport, healthcare and communications. Here we focus on climate change as a potential new area of application, specifically on energy efficiency, for two reasons: first, it is one of the greatest challenges facing mankind; second, the binary nature of the choices involved–i.e. energy-efficient versus energy-inefficient behaviour–makes climate change a perfectly suited field of application.

After decades of intense scientific debate, it has now been demonstrated beyond reasonable doubt that climate change and global warming are indeed taking place. It is almost unanimously accepted that human greenhouse gas emissions play a key role (Stern 2006). Given the current and projected levels of greenhouse gas emissions, climate models predict temperatures to rise significantly over the course of the next century. If nothing is done to reduce the emissions, serious consequences are predicted for our planet, including mass extinctions, sea level rises, and increase in the occurrences of extreme weather events such as hurricanes, flooding and severe drought (IPCC's Fouth Assessment Report 2007).

In recent years, due to a growing popular awareness, the debate has moved to the top of the political agenda. No country denies the existence of the problem, and many are already taking action to reduce emissions.

The famous Kyoto agreement in 1997 resulted in a set of emissions targets for a number of developed countries. Since this agreement is set to expire in 2012, governments have been actively working to reach a new agreement for the post-Kyoto framework. The UN Conference on Climate Change that took place in December 2007 in Bali resulted in a 'roadmap', whereby the international community agreed to begin negotiations towards a new global deal on climate change. Many countries are already taking unilateral action. For example, the United Kingdom is about to introduce a new Climate Change Bill, which will commit the nation to a legally binding target of emissions reductions.

Once the international post-Kyoto emission reduction targets are agreed, the question will be how to meet them. Figure 4 shows the various available options for reducing greenhouse gas emissions. Reducing the global population is not an option, although working on reducing population growth may be. Alternatively, we can reduce emissions per capita. This could be done by reducing our level of consumption. Although this clearly is one of the causes of the problem, it is unlikely that this option would produce significant emissions reductions in the short term. This is partly because it would require the population of the world to significantly change their current way of life, as well as to change the existing global economic system. Another alternative is to improve the carbon intensity of our energy sources–for example, by replacing coal power plants with wind farms. Much work is being done in this direction.

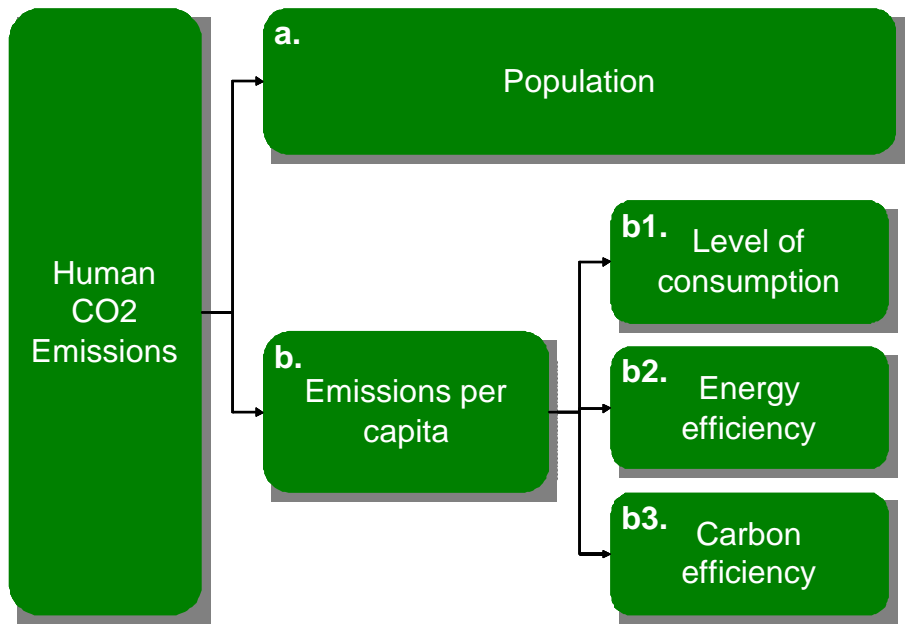

**Figure 4:** *All the ways to reduce greenhouse gas emissions.*

The last option is to increase energy efficiency. This option has a number of significant advantages, for example, it can potentially lead to significant financial savings. Some models estimate that an annual US$500bn could be globally by 2030, US$90bn of which in the USA alone (Creyts et al 2007). This would result from, for example, lower energy bills after thermally insulating homes or switching to energy efficient lighting. Moreover, higher energy efficiency would reduce the stress on the other options discussed above: for example, if no extra energy is required, then there is no need to build a wind farm instead of a coal power plant.

The models presented in this paper, together with the framework for deconstructing this apparently unmanageable problem into smaller manageable chunks, offer a systematic, robust and transparent approach to tackling the problem of climate change. Moreover, from a research perspective, climate change offers a testing ground to further develop and improve this family of models that can be applied to problems in most policy areas.

**3.1 Case study – United Kingdom**

Governments have generally adopted top-down policies to increase energy efficiency which are focused on particular actions and technologies. For example, since condensing boilers became mandatory in the UK in 2005, energy regulators require that suppliers undertake a pre-agreed level of activity to improve energy efficiency and save emissions in the domestic sector; the scheme is known as Carbon Emissions Reduction Target (DEFRA 2008). Neither Governments nor energy companies have yet engaged in incentivising individuals to change their behaviour. In fact, energy bill structures still reward customers who use more energy by offering them a lower per unit tariff. Regulating companies to undertake energy efficiency activities has the effect of subsidising these activities and potentially undermining any attempts to develop a profit driven market in these areas. Customers who know that energy companies are obliged to offer insulation, energy efficiency appliances and energy efficient light bulbs may be less inclined to pay for these things themselves.

The current set of regulations in the domestic sector is set to expire in 2011, and some energy companies (including Scottish & Southern Energy, the second largest supplier of electricity and natural gas in the UK) are demanding a less prescriptive regulatory approach. This means that they are asking for a legally binding target from the Government to reduce customers demand that incorporates the freedom to meet that target it in the most efficient way. A similar policy is expected to be introduced in the large commercial sector in 2010–the Carbon Reduction Commitment–which caps participants emissions from their downstream use of energy.

In summary, it is likely that new policies could be introduced in the UK to create a *new market in energy efficiency*. In this case large energy providers–and potentially new start-up companies–will be looking for ways to induce large segments of the population to adopt more energy efficient behaviours in order to reduce their demand.

To put this in the context of the models presented in this paper: practically, decision-makers want to know, for instance, what Government should do to induce the population to buy energy saving light bulbs. [The government could invest toward lowering the cost of each light bulb, or educating towards energy saving lifestyles; on the other hand, money could be saved by reducing $CO_2$ emissions, for example by generating carbon credits to be sold on the international carbon market, or avoiding penalties for non-compliance to international agreements such as the Kyoto Protocol.] How to balance the choice crucially depends on predicting the percentage of people who will turn to energy-saving light bulbs. That is what a statistical mechanics model can achieve after a suitable estimation of the parameters involved which include, in particular, the measure of what the imitation strength is between peers concerning buying habits.

More concretely, the private sector will play a major role in tackling climate change, and companies are already looking for ways to contribute to the solution as well as to make significant profits. For example, Philips is developing and delivering energy-efficient light bulbs to the market. An achievable energy saving of up to 40 percent on all the lighting currently installed globally would save 106 billion Euros. This equates to 555 million tonnes of $CO_2$ per year, which corresponds to 1.5 billion barrels of oil per year or the annual output of 530 medium sized power stations producing 2TWh per year (Verhaar 2007).

There is, therefore, an opportunity for governments and businesses to work together towards inducing sustainable behaviour by the consumer. The family of models presented in this paper offers a tool to provide evidence and inform decision-makers and help them make the relevant choices.

## 4. Beyond climate change: Why focus on behaviour and cultural change to achieve policy goals?

Most public and social policies are based on theoretical assumptions about human behaviour. However, these are rarely made explicit or tested against the available data. There are a number of factors that have encouraged the growing academic and policy interest in how to induce behaviour change amongst a population in order to generate sustainable and cost effective social improvements.

Delivering and achieving major sustainable policy outcomes on issues such as climate change requires greater engagement and participation from a national population than traditional ways of delivering public services or policies – *'you can't leave it all up to the government'*. Higher levels of spending and better-run public services can achieve improved outcomes. However, in order to achieve sustainable lasting outcomes and social improvements, much depends on changes in individual personal behaviour: for example in achieving population improvements in health and well-being. This means individuals adopting a better diet and taking up more exercise, and in education the emphasis is on children's willingness to learn and parents' willingness to help them learn (Knott *et al* 2007)

There are also strong moral and political arguments for encouraging personal responsibility and behavioural change amongst a population. Most of the dominant traditions of social and political thought emphasise individuals' and communities' ability to take control and act in their own best interests. They see it as better for governments to empower citizens and provide a social and economic context in which citizens are able to make informed decisions regarding their own behaviour.

And lastly, policy interventions based on behaviour change can be significantly more cost-effective and preventive than traditional service delivery. There is evidence from across a range of policy areas–for example in health, education, crime–of the cost-effectiveness of behaviour based social interventions. For example altering an individual's diet to one that reduces and prevents the risk of cardiovascular disease is cheaper and more efficient than dealing with the consequences of poor diet with heart surgery.

## 5. Conclusion

Climate change is now hovering near the national and global political agenda. Motivated by increasing public awareness, pressure from environmental organizations and a growing body of scientific evidence, decision-makers are now working hard to reach an international deal to reduce greenhouse gas emissions and avoid the devastating social, economic and environmental impacts of climate change. Emissions targets are expected to become increasingly strict over the next decades. Consequently, this means that governments and companies will look for the most cost-effective ways of meeting targets by providing the social and economic context in which people are able and willing to make informed choices regarding their lifestyles.

This paper has argued that increasing energy efficiency has a number of advantages, including potential annual savings worth hundreds of billions of dollars by 2030, and will play a key role in the future. It is expected that new regulations will allow the creation of new markets in energy efficiency, whereby profits would accrue by incentivising consumers to adopt energy efficient behaviours. There is already demand from energy companies, such as Scottish and Southern Energy in the UK, for such policies. Moreover, companies such as Philips would profit from selling new energy efficient bulbs.

This means that there is a growing demand for behavioural models that can help contribute to better understanding, in concrete and measurable ways, the drivers behind consumer choices. In particular, there is interest in models that may help policy-makers trigger structural changes in the way people behave. Given the nature of the problem of climate change, this demand is set to grow dramatically in the next few years.

This paper presented a family of models that can address this issue. These models combine the practicality and reputation of well-established econometric tools with the flexibility and rigor of advanced mathematical tools produced by decades of research in the physical sciences. Perhaps more importantly, these models will bring together the insights from experts in both the physical and social sciences.

Moreover, the models presented in this paper are consistent with the concept of personal responsibility, and may be used to empower individuals to make choices that are consistent with their personal interests as well as with the common good. Perhaps these models may help governments create a society that spontaneously protects the private as well as the public good in a way that reduces government interference but avoids market failures and 'tragedy of the commons' scenarios (Schelling 1978).

## Acknowledgements

F.G. and A.C. would like to thank Bryony Worthington and James Fox for their contributions and the useful discussions. I.G. acknowledges partial support from the CULTAPTATION project of the European Commission (FP6-2004-NEST-PATH-043434).

## References

AKERLOF, G. (1997) "Social distance and economic decisions", *Econometrica* 65, pp. 1005–1027.
ARIELY, D. (2008) "Predictably irrational – the hidden forces that shape our decisions", Harper Collins Publishers, London.
BANDURA, A. (1986) Social Foundations of Thought and Action: A Social Cognitive Theory. Englewood Cliffs, NJ: *Prentice Hall*.
BEN-AKIVA, M. and LERMAN, S. R. (1985), "Discrete Choice Analysis", The MIT Press, Cambridge, Mass.
BOND, R. and SMITH, P. B. (1996), "Culture and conformity: A meta-analysis of studies using Asch's (1952b,1956) line judgment task". *Psychological Bulletin* 119, 111–137.
BROCK, W. and DURLAUF, S. (2001), "Discrete Choice with Social Interactions", *Review of Economic Studies*, 68: 235-260.
BROCK, W. and DURLAUF, S. (2007), "Identification of Binary Choice Models with Social Interactions", *Journal of Econometrics*, 140: 52-75.
CONTUCCI, P., GALLO I. and GHIRLANDA, S. (2007) "Equilibria of culture contact derived from ingroup and outgroup attitudes", *arXiv: 0712.1119*.
CONTUCCI, P. and GHIRLANDA, S. (2007) "Modeling society with statistical mechanics: an application to cultural contact and immigration", *Quality and Quantity*.
CONTUCCI, P. and GIARDINA', C., (2008) "Mathematics and Social Sciences: A Statistical Mechanics Approach to Immigration", *ERCIM News* 73: 34-35.
CREYTS, J., DERKACH A., NYQUIST S., OSTROWSKI K. and STEPHENSON J. (2007) "Reducing U.S. greenhouse emissions: how much and at what cost?", *McKinsey&Company Report*, http://www.mckinsey.com/clientservice/ccsi/greenhousegas.asp
DEFRA (2008) "Household energy supplier obligations – The Carbon Emissions Reduction Target" Published on the DEFRA website:
http://www.defra.gov.uk/environment/climatechange/uk/household/supplier/index.htm
FOX, J., DALY, A.J. and GUNN, H. (2003) *"Review of RAND Europe's transport demand model systems"*. Published on RAND's website: http://rand.org/pubs/monograph_reports/MR1694
GERARD, K., SHANAHAN, M. and LOUVIERE, J. (2003) "Using stated Preference Discrete Choice Modelling to inform health care decision-making: A pilot study of breast screening participation", *Applied Economics*, 35, (9): 1073-1085.
GRANOVETTER, M. (1978), "Threshold Models of Collective Behavior", *The American Journal of Sociology,* 83: 1420-1443.

IDA, T. and KURODA, T. (2006) "Discrete choice analysis of demand for broadband in Japan" *Journal of Regulatory Economics*, 29, (1): 5-22

IPCC (2007) *"Fourth assessment report: climate change 2007"* Published on the IPCC's website: http://www.ipcc.ch/ipccreports/assessments-reports.htm

KNOTT, D. MUERS, S. and ALDRIGE, S (2007) "Achieving Cultural Change:A Policy Framework". The Strategy Unit, Cabinet Office, UK Government.

LUCE, R. and SUPPES, P. (1965), "Preferences, Utility and Subjective Probability", in *Handbook of Mathematical Psychology,* Vol. 3, Luce R., Bush R. and Galenter E., eds. Wiley, New York.

MCFADDEN, D. (2001), "Economic Choices", *The American Economic Review*, 91: 351-378.

ORTUZAR, J. and WILUMSEN, L. (2001) *"Modelling Transport"*. Wiley, Chichester, UK.

PAAG, H., DALY, A.J., ROHR, C. (2001) *"Predicting use of the Copenhagen harbour tunnel"*. In *"Travel behaviour research: the leading edge"* David Hensher (ed.), Pergamon.

PERSKY J. (1995) 'Retrospectives: The Ethology of Homo Economicus', *The Journal of Economic Perspectives*, Vol. 9, No. 2, pp. 221-23

RYAN, M and GERARD, K (2003) "Using discrete choice experiments to value health care programmes: current practice and future research reflections", *Applied Health Economics and Health Policy*, 2, (1): 55-64.

RYAN, M., NETTEN A., SKATUN D. and SMITH P (2006) "Using discrete choice experiments to estimate a preference-based measure of outcome – An application to social care for older people" *Journal of Health Economics*, 25, (5): 927-944.

SCHEINKMAN, J. A. (2008), "Social interactions", *The New Palgrave Dictionary of Economics*, 2nd Edition, Palgrave Macmillan.

SCHELLING, T. (1978), "Micromotives and Macrobehavior", *W. W. Norton & Company*, New York.

STERN, N. (2007) "The economics of climate change – The Stern Review", *Cambridge University Press*, Cambridge.

TRAIN, K. (2003) *"Discrete choice methods with simulation"*. Cambridge University Press, Cambridge, UK.

VERHAAR, H. (2007) "Reducing $CO_2$ emissions by 555 Mton through Energy Efficiency Lighting". Presented at the UNFCCC Conference in Bali, December 8.

WEISS, P. (1907) "L'hypothèse du champ moléculaire et la propriété ferromagnétique", *J. de Phys.*, 4 série, VI: 661-690